\documentclass[prl,aps,twocolumn,epsfig,superscriptaddress,showpacs]{revtex4}
\usepackage{mathrsfs}
\usepackage{amsfonts}
\usepackage{amsmath,graphicx,epsfig,color,CJK}
\usepackage{pifont,bbding,amssymb}
\usepackage[hypertex]{hyperref}

\begin{document}
\title{Bell Function Values Approach to Topological Quantum Phase Transitions}
\author{Dong-Ling Deng}
 \affiliation{Theoretical Physics
Division, Chern Institute of Mathematics, Nankai University, Tianjin
300071, People's Republic of China}
\affiliation{Centre for Quantum
Technologies and Department of Physics, National University of
Singapore, 117543, Singapore}

\author{Chunfeng Wu}
\affiliation{Centre for Quantum Technologies and Department of
Physics, National University of Singapore, 117543, Singapore}

\author{Jing-Ling Chen}
 \email{chenjl@nankai.edu.cn}
\affiliation{Theoretical Physics Division, Chern Institute of
Mathematics, Nankai University, Tianjin 300071, People's Republic of
China}

\author{Shi-Jian Gu}
\affiliation{Department of Physics and ITP, The Chinese University
of Hong Kong, Hong Kong, China}

\author{Sixia Yu}
\affiliation{Centre for Quantum Technologies and Department of
Physics, National University of Singapore, 117543, Singapore}

\author{C. H. Oh}
\email{phyohch@nus.edu.sg}
 \affiliation{Centre for Quantum
Technologies and Department of Physics, National University of
Singapore, 117543, Singapore}

\date{\today}

\begin{abstract}
We investigate the relation between Bell function values (BFV) of
the reduced density matrix and the topological quantum phase
transitions
in the Kitaev-Castelnovo-Chamon model. 
We find that the first order derivative of BFV exhibits singular
behavior at the critical point and we propose that it can serve as a
good and convenient marker for the transition point. More
interestingly, the value of the critical point can be analytically
obtained in this approach. Since the BFV serves as a measure of
nonlocality when it is greater than the classical bound of the
correlation functions, our work has established a link between
quantum nonlocality and phase transitions.

\end{abstract}


\maketitle

Topological phase of some strongly correlated quantum many body
systems is a new kind of order that depends on the system
topology~\cite{Wen-Book}. It has attracted great interest recently
because it can exhibit remarkable phenomena such as quasiparticles
with anyonic statistics. An archetypal physical realization of such
phase is in the quantum Hall system~\cite{1990Prange-Book}, which
bears many unconventional characteristics, such as fractional
statistical behaviors and ground state topological degeneracy that
cannot be lifted by any local perturbations~\cite{1985Haldane}. A
particular interest in topological ordered states is their
robustness against local perturbations which can lead to several
consequences such as topological insulators~\cite{2009Hsieh} and
topological quantum computations~\cite{2008Nayak}.

Not surprisingly, the unconventional properties of topological phase
might result in exotic critical phenomena, which cannot be
characterized by the Landau-Ginzburg-Wilson spontaneous
symmetry-breaking theory where the correlation function of local
order parameters plays a crucial role~\cite{1999Sachdev-Book}. For
example, the quantum phase transition between an Abelian and a
non-Abelian topological phase in chiral spin liquid might be
characterized by global flux and generalized topological
entanglement entropy~\cite{2009Chung}. More remarkably, for
time-reversal invariant anyonic quantum systems, Gils \textit{et
al}., have recently showed that the topological phases could be
uniformly described in terms of fluctuations of the two-dimensional
surfaces and their topological changes~\cite{2009Gils}. However, an
universal characterization and detection of topological phase and
its transitions still pose a big challenge despite a vast amount of
prominent works dealing with this problem.

During the past few years, several important concepts in the quantum
information field have been borrowed to characterize quantum phase
transitions (QPTs) and topological quantum phase transitions
(TQPTs), these including
entanglement~\cite{2002Osterloh-entanglement},
fidelity~\cite{2006Quan-Fidelity}, fidelity
susceptibility~\cite{2007You-FS}, and
discord~\cite{2008Dillenschneider-discord}, etc. A brief review of
the progress related to this issue is given in Ref.~\cite{2008Gu}
and references therein. Notwithstanding the great successes in
marking QPTs and TQPTs in some physical systems, each approach above
has its own disadvantages~\cite{2008Gu}. Take the fidelity approach
for example, to witness the QPTs, one has to find out the exact
ground state. However, for most of the physical systems, finding out
 the exact ground state is very difficult. In addition, it is also a
challenge to measure the fidelity in experiment on scalable systems.
An alternative choice is to use Bell function values (BFV) as
defined in expression (\ref{BFV-definition}) below, which indicates
the correlations of a quantum system and measures quantum
nonlocality when it is greater than the classical bound of the
correlation functions~\cite{2009Forster}. Actually, besides
entanglement, quantum nonlcoality is also a central nonintuitive
phenomena of quantum mechanics and it plays a key role in many
quantum information and computation processes, such as quantum key
distribution (KQD)~\cite{1991Ekert-Cryptograph}, nonlocal quantum
computation~\cite{2007Linden}, etc. 
Naturely, one would ask whether nonlocality can mark QPTs and TQPTs?

In this Letter, we propose the use of BFV as the marker of TQPTs and
provide a positive answer to the above question by investigating the
relation between BFV of the reduced density matrix and the TQPTs.
The motivation for choosing BFV is two-fold: (i) BFV can measure the
nonlocality of a quantum system, thus it might establish a
connection between nonlocality and TQPTs, which belong to two
different aspects; (ii) To get the BFV in an experiment scheme, one
only has to do some measurement on the qubits instead of knowing
exactly the ground state. Thus, our approach has its advantages in
experimental schemes. The discussion here is mainly based on the
Kitaev-Castelnovo-Chamon model~\cite{2008Castelnovo}, which exhibits
a second-order TQPT at the critical point. Our results indicate that
the first order derivative of BFV shows singular behavior at the
transition point. More interestingly, through this approach, one can
analytically obtain the critical value of the transition point.
Finally, using BFV to signal TQPTs and QPTs in other systems is also
briefly discussed.

%

\textit{Bell function values (BFV)}.---The famous Bell-
Clauser-Horne-Shimony-Holt (Bell-CHSH) inequality for two entangled
spin-$1/2$ (or qubit) particles, which has always provided an
excellent test-bed for experimental verification of quantum
mechanics against the predictions of local realism, is given by the
inequality~\cite{1964Bell}:
\begin{eqnarray}\label{CHSH-Ineq}
\mathcal{I}=Q_{11}+Q_{12}+Q_{21}-Q_{22}\leq2,
\end{eqnarray}
where $Q_{ij}=\int_\Gamma\mu(\lambda)
X_1(\mathbf{n}_i^{X_1},\lambda)X_2(\mathbf{n}_j^{X_2},\lambda)d\lambda$
is the correlation function with $X_k(\mathbf{n}_m^{X_k},\lambda)$
denoting the $m$-th observable on the $k$-th particle (here
$i,j,k,m=1,2$);
 $\Gamma$ is the total space of the hidden variable
$\lambda$ and $\mu(\lambda)$ is a statistical distribution of
$\lambda$, satisfying $\int_{\Gamma}\mu(\lambda)d\lambda=1$. Quantum
mechanically, the above inequality is violated by all pure entangled
states of two qubits~\cite{1991Gisin-theory} and the expression of
the correlation function for any two-qubit state $\rho$ reads:
$Q_{ij}^Q=\texttt{Tr}[(\mathbf{n}_i^{X_1}\cdot \vec{\sigma}) \otimes
(\mathbf{n}_j^{X_2}\cdot \vec{\sigma}) \rho]$. Here
$\mathbf{n}_m^{X_k}$ ($m,k=1,2$) are the unit vectors in
three-dimensional Hilbert space and $\vec{\sigma}$ is the Pauli
matrix vector. Based on the Bell-CHSH inequality~(\ref{CHSH-Ineq}),
the Bell function values is defined as:
\begin{eqnarray}\label{BFV-definition}
\mathscr{B}(\rho)\equiv\max \mathcal{I}^Q,
\end{eqnarray}
where $\mathcal{I}^Q=Q_{11}^Q+Q_{12}^Q+Q_{21}^Q-Q_{22}^Q$ and the
maximization is performed over all possible vectors
$\mathbf{n}_m^{X_k}$. Generally speaking, for every specific
two-qubit quantum state $\rho$, we need to carry out the procedure
of the maximization to obtain its BFV $\mathscr{B}(\rho)$.
Fortunately, in Ref.~\cite{1995Horodecki}, the authors introduced
another method to calculate $\mathscr{B}(\rho)$, which can
circumvent the tedious maximization.  It was proved there that
\begin{eqnarray}\label{BFVs-Horodecki}
\mathscr{B}(\rho)=2\sqrt{\upsilon_1+\upsilon_2},
\end{eqnarray}
where $\upsilon_1$ and $\upsilon_2$ are the two greater eigenvalues
of the $3\times3$ symmetric matrix $\mathscr
{L}_{\rho}^\texttt{T}\mathscr {L}_{\rho}$; $\mathscr {L}_{\rho}$ is
 a $3\times3$ matrix with elements defined by
$(\mathscr{L}_{\rho})_{\varsigma\tau}=\texttt{Tr}[\rho\sigma_{\varsigma}\otimes
\sigma_{\tau}]$ ($\varsigma,\tau=1,2,3$) and $\mathscr
{L}_{\rho}^\texttt{T}$ is the transpose of $\mathscr {L}_{\rho}$. In
the experimental situation, in order to obtain the BFV, the
observers of the first (second) qubit should carry out two
measurements $\mathbf{n}_1^{X_1}\cdot\vec{\sigma}$ and
$\mathbf{n}_2^{X_1}\cdot\vec{\sigma}$
($\mathbf{n}_1^{X_2}\cdot\vec{\sigma}$ and
$\mathbf{n}_2^{X_2}\cdot\vec{\sigma}$), just the same as in many
Bell-CHSH inequality testing experiments~\cite{1982Aspect}.

\begin{figure}
\includegraphics[width=75mm]{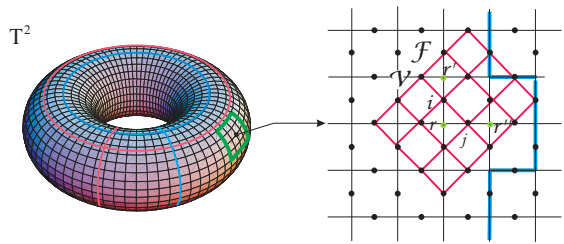}\\
 \caption{(Color online) An illustration of the Kitaev-Castelnovo-Chamon spin-lattice
model and its map to the $2$D Ising model: In the KCC model, each
black dot on the edge of the lattice represents a qubit and
$\mathcal{V}$ and $\mathcal{F}$ denote the vertex and the face,
respectively; The blue line across the lattices stands for a string
operator along the non-trivial loop on the torus $\texttt{T}^2$. The
system is invariant under transformation along the blue
line~\cite{2009Chen}. In its corresponding $2$D Ising model, the
qubits live on the vertex (green dots). Mapping:
$\sigma_i^z=\theta_r\theta_{r'}$, where $i$ is the bond between the
neighboring vertices $\langle r,r'\rangle$. Thus, for $i$ and $j$
nearest neighbors, the mapping gives $\langle
\sigma_i^z\sigma_j^z\rangle=\langle\theta_r\theta_{r'}\theta_{r''}\theta_r
\rangle=\langle\theta_{r'}\theta_{r''}\rangle$, namely, the nearest
qubits in KCC model become next-nearest in the corresponding $2$D
Ising model.}\label{fig1}
\end{figure}

\textit{The Kitaev-Castelnovo-Chamon (KCC) Model}.---The physical
model we consider in this article was introduced by Castelnovo and
Chamon~\cite{2008Castelnovo}, which is a deformation of the Kitaev
toric code model~\cite{2003Kitaev}. The Hamiltonian of the KCC model
with periodic boundary conditions reads:
\begin{eqnarray}
\mathscr{H}=-\mathcal{J}_m\sum_{\mathcal{F}\in
\texttt{T}^2}B_{\mathcal{F}}-\mathcal{J}_e\sum_{\mathcal{V}\in
\texttt{T}^2}A_\mathcal{V}+\mathcal{J}_e\sum_{\mathcal{V}\in
\texttt{T}^2}e^{-\beta\sum_{j\in \mathcal{V}}\sigma^z_j},\nonumber
\end{eqnarray}
where $\mathcal{J}_m,\mathcal{J}_e>0$, $\beta$ is a coupling
constant; $A_\mathcal{V}=\prod_{j\in \mathcal{V}}\sigma_j^x$ and
$B_\mathcal{F}=\prod_{j\in \mathcal{F}}\sigma_j^z$ are the vertex
and face operators in the original Kitaev toric code
model~\cite{2003Kitaev}. A brief sketch of this model is shown in
Fig. \ref{fig1}. The ground state in the topological sector
containing the fully magnetized state
$|0\rangle=|\uparrow\uparrow\cdots\uparrow\rangle$ can be
analytically obtained~\cite{2008Castelnovo}:
\begin{eqnarray}
|G(\beta)\rangle=\mathcal{Z}(\beta)^{-\frac{1}{2}}
\sum_{g\in\mathcal{G}}e^{\beta\sum_j\sigma_j^z(g)/2}g|0\rangle,
\end{eqnarray}
where $\mathcal{Z}(\beta)=\sum_{g\in\mathcal{G}}
e^{\beta\sum_j\sigma_j^z(g)}$; $\mathcal{G}$ is the Abelian group
generated by the vertex operators $\{A_{\mathcal{V}}\}$ and
$\sigma_j^z(g)$ is the value of spin at site $j$ in state
$g|0\rangle$. Obviously, when $\beta=0$, $|G(\beta)\rangle$ reduces
to the topologically ordered ground state of the toric code
model~\cite{2003Kitaev}, while when $\beta\rightarrow\infty$,
$|G(\beta)\rangle$ becomes the fully magnetized state $|0\rangle$.
At the point $\beta_c=\frac{1}{2}\ln(\sqrt{2}+1)$ there exists a
second-order TQPT where the topological entanglement entropy
$\mathcal{S}_{topo}=1$ for $\beta<\beta_c$ changes to
$\mathcal{S}_{topo}=0$ for $\beta>\beta_c$~\cite{2008Castelnovo}.

As shown by Castelnovo and Chamon, there exists a one-to-two mapping
between the configurations $\{g\}=\mathcal {G}$ the configurations
$\{\theta\}$ of the classical $2$D Ising
model~\cite{2008Castelnovo}. In the mapping, the Hamiltonian of the
Ising model has the form $\mathscr{H}_{Ising}=-\mathscr{C}
\sum_{\langle r,r'\rangle}\theta_r\theta_{r'}$, where $\mathscr {C}$
is a coupling constant and $\theta_r,\theta_{r'}=+1$ or $-1$
depending on whether or not the corresponding vertex operator
$A_{\mathcal{V}}$ is acting on the site $r$. Thus
$\sigma^z_i=\theta_r\theta_{r'}$ with $i$ being the edge between the
nearest neighboring vertexes. An illustration of this mapping is
 shown in Fig.~\ref{fig1}.

\begin{figure}
\includegraphics[width=87mm]{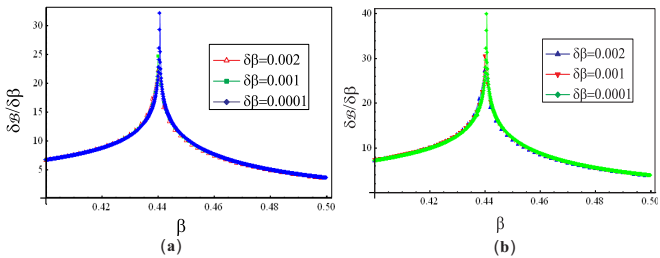}\\
 \caption{(Color online) Numerical results of the first order derivative of
 BFV $\mathscr{B}(\rho_{ij})$ versus $\beta$ for different $\delta\beta$. (a) $i$ and $j$
 are the nearest case; (b) $i$ and $j$
 are the next-to-nearest case.}\label{fig2}
\end{figure}

\textit{Signaling TQPTs by BFV}.---Since the BFV introduced in
expression~(\ref{BFV-definition}) only account for two-qubit states,
we need to calculate the reduced density matrix of two qubit
$\rho_{ij}$ based on the ground state $|G(\beta)\rangle$ and the
symmetry of the Hamiltonian $\mathscr{H}$. It was shown in
Ref.~\cite{2009Eriksson} that $\rho_{ij}$ has the following form
(details are given in Ref.~\cite{2009Chen,2008Castelnovo} and
references there in):
\begin{eqnarray}\label{Reduced-Matrix}
\rho_{ij}=\frac{1}{4}[\mathbf{I}+\langle
\sigma_i^z\rangle(\sigma^z_i+\sigma^z_j)+\langle\sigma_i^z\sigma_j^z\rangle
\sigma_i^z\sigma^z_j],
\end{eqnarray}
where $\mathbf{I}$ is the $4\times4$ identity matrix. Based on the
Eq.~(\ref{Reduced-Matrix}), the BFV can be calculated by using the
simplified formula for $\mathscr{B}(\rho_{ij})$ in
Eq.~(\ref{BFVs-Horodecki}). For convenience and simplicity, we only
concentrate on two cases where $i$ and $j$ are nearest and
next-to-nearest neighbors, respectively. In the thermodynamic limit,
the mapping to the $2$D Ising model gives that
$\langle\sigma^z_i\rangle=\langle\theta_r\theta_{r'}\rangle=
-\coth(2\beta)[\pi+(4\tanh^2(2\beta)-2)\mathscr{X}(\chi)]/(2\pi)$,
where $\mathscr{X}(\chi)=\int_0^{\pi/2}d \varphi
(1-\chi^2\sin^2\varphi)^{-1/2}$ and
$\chi=2\sinh(2\beta)/\cosh^2(2\beta)$. For the calculation of
$\langle\sigma_i^z\sigma_j^z\rangle$, the equivalence between the
$2$D Ising model and the quantum $1$D XY model yields:

(i) For $i$ and $j$ the nearest case,
$\langle\sigma_i^z\sigma_j^z\rangle=\langle\theta_r\theta_{r'}\rangle
=\frac{1}{\pi}\int_0^{\pi}d\phi\{[\sinh^{-2}(2\beta)-\cos{\phi}]\cos
{\phi}-\sin^2{\phi}\}/\{
\sin^2{\phi}+[\sinh^{-2}(2\beta)-\cos{\phi}]^2\}^{1/2}$. %
Summarizing all the relations above enables us to obtain the BFV
$\mathscr{B}(\rho_{ij})$. For the $i$ and $j$ nearest case, the
numerical results for the first order derivative of BFV
$\frac{d\mathscr{B}(\rho_{ij})}{d\beta}$ are displayed in
Fig.~\ref{fig2}(a), from which we see a distinct rapid increase of
$\frac{d\mathscr{B}(\rho_{ij})}{d\beta}$ around the critical point
$\beta_c\approx0.44$. Note that we only focus on a small neighboring
region of $\beta$ around the TQPT point $\beta_c$, namely
$0.4\leq\beta\leq0.5$. The smaller the $\delta \beta$, the greater
the rapid increase. When $\delta\beta\rightarrow0$,
$\frac{d\mathscr{B}(\rho_{ij})}{d\beta}\rightarrow +\infty$,
indicating its singularity at the TQPT point $\beta_c$.

%

More interestingly, after long tedious but straightforward
calculations, we arrive at an analytical formula for
$\frac{d\mathscr{B}(\rho_{ij})}{d\beta}$, which can enable us to
obtain the analytical value of $\beta_c$:
\begin{eqnarray}\label{FirtDr-Analytic}
\frac{d\mathscr{B}}{d \beta}=\int_0^{\pi}d\phi
\texttt{csch}^2(2\beta)\sin^2\phi\Upsilon(\phi,\beta)/\pi,
\end{eqnarray}
Here $\Upsilon(\phi,\beta)= 8\coth{(2\beta)} \texttt{csch}^2(2
\beta)/(1-2\cos\phi\texttt{csch}^2(2\beta)+\texttt{csch}^4(2\beta))^{3/2}$.
What is interesting is that we can analytically obtain the critical
point from the Eq. (\ref{FirtDr-Analytic}). To this end, one can
rewrite $\Upsilon(\phi,\beta)$ as
$\Upsilon(\phi,\beta)=2\sqrt{2}\coth(2\beta)/\{[\frac{1}{2}(\texttt{csch}^2(2\beta)
+1/\texttt{csch}^2(2\beta))-\cos\phi]^{3/2}\texttt{csch}(2\beta)\}$.
Obviously, Eq. (\ref{FirtDr-Analytic}) has only one singular point
because $\texttt{csch}^2(2\beta) +1/\texttt{csch}^2(2\beta)\geq2$
and so the singularity happens at $\texttt{csch}^2(2\beta)
+1/\texttt{csch}^2(2\beta)=2$, namely
$\beta_c=\frac{1}{2}\ln(\sqrt{2}+1)$. This explicitly exhibits one
of the advantages of BFV approach to TQPTs.

For $i$ and $j$ the next-to-nearest case, direct calculations show
that $\langle\sigma_i^z\sigma_j^z\rangle=\cosh^2(\beta^*)
(\mathcal{T}_{-1}^2-\mathcal{T}_{-2}\mathcal{T}_0)-
\sinh^2(\beta^*)(\mathcal{T}_{1}^2-\mathcal{T}_{2}\mathcal{T}_0)$,
where
$\mathcal{T}_{\kappa}=\frac{1}{\pi}\int_0^{\pi}d\phi[(\xi-\cos{\phi})\cos
{\kappa\phi}+\gamma\sin{\phi}\sin{\kappa\phi}]/[
(\gamma\sin{\phi})^2+(\xi-\cos{\phi})^2]^{1/2}$. Here
$\tanh(\beta^*)=e^{-2\beta}$, $\gamma=[\cosh(2\beta^*)]^{-1}$ and
$\xi=(1-\gamma^2)^{1/2}/\tanh(2\beta)$. We also plot the numerical
results of $\frac{d\mathscr{B}(\rho_{ij})}{d\beta}$ in
Fig.~\ref{fig2}(b). From this figure, one can observe that the
$\frac{d\mathscr{B}(\rho_{ij})}{d\beta}$ has a singularity at the
TQPT point $\beta_c$. It is also obvious that
$\frac{d\mathscr{B}(\rho_{ij})}{d\beta}$ behaves quite similarly
between the nearest and next-to-nearest case. 
This result accords with the results in Ref.~\cite{2009Eriksson},
where the reduced fidelity and reduced fidelity susceptibilities are
only slightly different between the two cases, respectively. 
Since BFV indicates the qubit correlations in the system, it seems
that the correlations of nearest qubits is similar as that of
next-to-nearest qubits in topological ordered states. This is, to
some extent, counterintuitive because in many physical systems the
interaction between nearest particles is usually greater than that
of next nearest particles.

It is worthwhile to note that the reduced density matrix $\rho_{ij}$
is diagonal. That indicates the correlations between any two local
spins in the ground state of KCC model are always classical.
Consequently, the Bell function values of $\rho_{ij}$ cannot be
greater than $2$, the classical bound. Another interesting
consideration here is similar to Ref.~\cite{2009Chen}: we can
calculate the BFV between a local qubit denoted by $i$ and the rest
of the whole lattice by rewriting the ground state as
$|\texttt{G}(\beta)\rangle=\mathscr{Y}_+|\mathscr{P}\rangle|0\rangle_i+
\mathscr{Y}_-|\mathscr{Q}\rangle|1\rangle_i$, where
$\mathscr{Y}^2_{\pm}=(1\pm\langle\theta_{0,0}\theta_{0,1}\rangle)/2$,
$|\mathscr{P}\rangle$ and $|\mathscr{Q}\rangle$ are two orthogonal
normalized vectors. Consequently, we can regard
$|\texttt{G}(\beta)\rangle$ as a simple pure two-qubit entangled
state. In this case, the BFV has a one-to-one monotonous relation
with entanglement~\cite{1992Popescu}, thus also with quantum discord
since for pure two-qubit state, the quantum discord is the same as
entanglement of entropy~\cite{2003Vedral}. As a result, the BFV
should behave similarly as the quantum discord does at the critical
point $\beta_c$ (for details, see Ref.~\cite{2009Chen}). However, it
is worthwhile to clarify that there is a distinctive difference
between the BFV approach and the quantum discord approach. For the
quantum discord, its value becomes trivially $0$ for the reduced
two-qubit state, thus cannot signal the TQPT at the critical point.
Nevertheless, as shown in the former paragraphs, the first order
derivative of BFV is an excellent marker of the transitions. In
addition, the physical meanings of BFV and quantum discord are
different. Generally speaking, quantum discord is a measurement of
the \textit{`quantumness'} of a system. While, BFV measures the
nonlocality of the system when it is greater than the classical
bound. In this case, $\mathscr{B}(|\texttt{G}(\beta)\rangle)
=2\sqrt{1+4\mathscr{Y}^2_+\mathscr{Y}^2_-}>2$. Thus the BFV
$\mathscr{B}(|\texttt{G}(\beta)\rangle)$ can measure the nonlocality
of ground state. This establishes a new link between quantum
nonlocality and TQPTs.

%

\textit{Summary and Discussion}.---To summarize, based on the KCC
model, which exhibits a second-order TQPT at the critical point, we
have introduced the BFV approach to TQPTs. Our results show that BFV
serves as an accurate marker of the transitions. Since the BFV also
serves as a measure of nonlocality, which is a pure quantum
phenomenon and cannot be described by any local realism theory, our
work has established a new link between quantum nonlocality and
phase transitions. Furthermore, experimentally, this approach only
involves two measurements on two qubit, thus it might be more
convenient to implement in experimental schemes. Actually, the
optical lattices and trapped ions might provide suitable
experimental test-bed for our results~\cite{2003Duan}.

What is also notable is that this approach is applicable to other
models. For instance, for the model recently introduced by Son
\textit{et al}., which is described by a cluster Hamiltonian
$\mathscr{H}(\lambda)=-\sum_{i=1}^{N}(\sigma_{i-1}^x\sigma_i^z\sigma^x_{i+1}+
\lambda\sigma_i^y\sigma_{i+1}^y)$ and has an exotic phase transition
at the critical point $\lambda=1$~\cite{2010Son}, our numerical
results show that the first order derivative of BFV can explicitly
capture the transition. The cluster Hamiltonian above can be
simulated in a triangular configuration of an optical lattice of two
atomic species~\cite{2010Son}, thus also leading to the possibility
of testing the BFV approach experimentally. To investigate the BFV
in QPTs, we have also considered the one dimensional Ising model and
XY model. The numerical results show that the first order derivative
of BFV exhibits singular behavior at the critical point, too. Thus
this approach is useful for both QPTs and TQPTs.

It would be interesting and significant to apply this approach to
QPTs and TQPTs in various physical systems, such as quantum spin
Hall system, both theoretically and experimentally. It would also be
interesting to use BFV based on other Bell inequalities to
investigate QPTs and TQPTs. More specifically, studying the BFV of
the pure ground states  based on multipartite Bell inequalities,
such as the famous Mermin-Ardehali-Belinskii-Klyshko (MABK)
inequality~\cite{MABK-Inequality}, might shed light on the behavior
of the quantum nonlocality of the whole system in QPTs and TQPTs.

The work is supported in part by National Research Foundation and
Ministry of Education, Singapore under research grant No. WBS:
R-710-000-008-271, in part by NSF of China (Grant No. 10975075),
Program for New Century Excellent Talents in University, and the
Project-sponsored by SRF for ROCS, SEM, and in part by the Earmarked
Grant Research from the Research Grants Council of HKSAR, China
(Project No. HKUST3/CRF/09).

\end{document}